\documentclass[twocolumn,showpacs,preprintnumbers,amsmath,amssymb, superscriptaddress, prb]{revtex4-1}

\usepackage{graphicx}

\usepackage{textcomp}

\usepackage{color}

\definecolor{red}{rgb}{1,0,0}

\definecolor{green}{rgb}{0,1,0}

\definecolor{blue}{rgb}{0,0,1}

\newcommand{\beq}{\begin{equation}}

\newcommand{\eneq}{\end{equation}} 

\newcommand{\beqa}{\begin{eqnarray}}

\newcommand{\eneqa}{\end{eqnarray}} 

\newcommand{\nn}{\nonumber} 

\newcommand{\bta}{\begin{tabular}}

\newcommand{\enta}{\end{tabular}}

\begin{document}

\title{Multiferroic FeTe$_2$O$_5$Br:
Alternating spin chains with frustrated interchain interactions}

\author{M. Pregelj}

\affiliation{Institute "Jo\v{z}ef Stefan", Jamova c.\ 39, SI-1000 Ljubljana, Slovenia}

\affiliation{Laboratory for Neutron Scattering, PSI, CH-5232 Villigen, Switzerland}

\author{H. O. Jeschke}

\affiliation{Institut f\"{u}r Theoretische Physik, Goethe-Uni\-ver\-si\-t\"at Frank\-furt am Main, 60438 Frankfurt am Main, Germany}

\author{H. Feldner}

\affiliation{Institut f\"{u}r Theoretische Physik, Goethe-Uni\-ver\-si\-t\"at Frank\-furt am Main, 60438 Frankfurt am Main, Germany}

\author{R. Valent\'{i}}

\affiliation{Institut f\"{u}r Theoretische Physik, Goethe-Uni\-ver\-si\-t\"at Frank\-furt am Main, 60438 Frankfurt am Main, Germany}

\author{A. Honecker}

\affiliation{Institut f\"ur Theoretische Physik and Fakult\"at f\"ur Mathematik und Informatik,
        Georg-August-Universit\"at G\"ottingen, Germany}

\author{T. Saha-Dasgupta}

\affiliation{Satyandranath Bose National Centre for Basic Sciences, Kolkata 700098, India}

\author{H. Das}

\affiliation{Satyandranath Bose National Centre for Basic Sciences, Kolkata 700098, India}

\author{S. Yoshii}

\affiliation{Institute for Materials Research, Tohoku University, Sendai 980-8577, Japan}

\author{T. Morioka}

\affiliation{Institute for Materials Research, Tohoku University, Sendai 980-8577, Japan}

\author{H. Nojiri}

\affiliation{Institute for Materials Research, Tohoku University, Sendai 980-8577, Japan}

\author{H. Berger}

\affiliation{\'{E}cole Polytechnique F\'{e}d\'{e}rale de Lausanne, Switzerland}

\author{A. Zorko}

\affiliation{Institute "Jo\v{z}ef Stefan", Jamova c.\ 39, SI-1000 Ljubljana, Slovenia}

\affiliation{EN--FIST Centre of Excellence, Dunajska 156, SI-1000 Ljubljana, Slovenia}

\author{O. Zaharko}

\affiliation{Laboratory for Neutron Scattering, PSI, CH-5232 Villigen, Switzerland}

\author{D. Ar\v{c}on}

\affiliation{Institute "Jo\v{z}ef Stefan", Jamova c.\ 39, SI-1000 Ljubljana, Slovenia}

\affiliation{Faculty of mathematics and physics, University of Ljubljana, Jadranska c.\ 19, SI-1000 Ljubljana, Slovenia}

\date{\today}

\begin{abstract}

A combination of density functional theory calculations, many-body model 
considerations, magnetization and electron spin resonance measurements 
shows that the multiferroic FeTe$_2$O$_5$Br should be described as a 
system of alternating antiferromagnetic $S=5/2$ chains with strong 
Fe-O-Te-O-Fe bridges weakly coupled by two-dimensional frustrated 
interactions, rather than the previously reported tetramer models. The 
peculiar temperature dependence of the incommensurate magnetic vector can 
be explained in terms of interchain exchange striction being responsible 
for the emergent net electric polarization.

\end{abstract}


\maketitle

\section{Introduction}

Frustrated low-dimensional spin systems exhibit a plethora of exotic magnetic ground
states, which are an exciting challenge both for theorists and experimentalists.\cite{Lacroix} 
Since quantum fluctuations enhanced by frustration tend to destabilize classically ordered states, such systems  often develop complex orders with broken inversion symmetry thus fulfilling the fundamental condition for multiferroicity.
This can lead to strong magnetoelectric (ME) coupling,\cite{Cheong, Khomskii, Eerenstein} most spectacularly observed as reversal of the electric polarization with the magnetic field.\cite{Hur} So far, strong ME coupling has been almost exclusively associated with transition metal (TM) oxides.

Unconventional magnetic ground states and multiferroicity
have been recently found also in TM selenium compounds \cite{Lawes,Bos}
and tellurite oxohalides.\cite{Zaharko,PregeljPRL, Hiroshi}
The peculiarity of these compounds lies within their exchange network, because TM ions are
coupled through complex pathways that 
involve several atoms, including tellurium \cite{Becker, Becker2} and selenium.\cite{Janson, Janson2}
This makes determination of the exchange pathways much less intuitive \cite{Deisenhofer,Das,Valenti} as opposed to the TM oxides, where dominant TM-O-TM exchange interactions can be
guessed from the bonding angles.\cite{Goodenough, Kanamori}
The structural complexity thus impedes the understanding of the microscopic picture of the exchange interactions and hence blurs the ME coupling mechanism, which both are yet to be established for this rapidly growing class of materials.

\begin{figure} [!]

\includegraphics[width=0.48\textwidth]{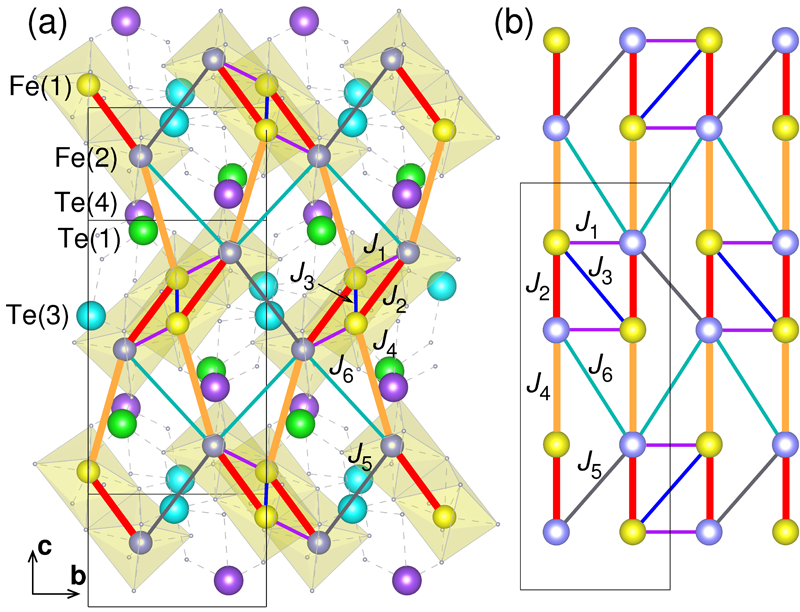}

\caption{(Color online) (a) Crystal structure of FeTe$_2$O$_5$Br along the $bc$ plane
where light yellow polyhedra denote [Fe$_4$O$_{16}$]$^{20-}$ tetramer units.
For clarity only Fe and Te atoms are shown.
(b) The DFT magnetic exchange network, where the strongest $J_2$ and $J_4$ (thick lines) form alternating spin chains coupled by weaker frustrated interactions (thin lines).
The marked unit cell contains 8 magnetic Fe$^{3+}$ ions. }
\label{figDFT}

\end{figure}

As a prominent example
we refer here to FeTe$_2$O$_5$Br,\cite{Becker, PregeljPRL} which adopts a layered structure
of [Fe$_4$O$_{16}$]$^{20-}$ tetramer 
clusters connected via Te$^{4+}$ ions [Fig.~\ref{figDFT}(a)]. The negative Curie-Weiss temperature 
 $T_{CW}$ = $-$98 K implies strong antiferromagnetic (AFM) interactions
 between the  Fe$^{3+}$ ($S$\,=\,5/2) moments, while the system develops
 long-range magnetic order at a 
considerably lower temperature $T_{N1}$\,=\,11\,K.\cite{Becker,PregeljPRL,PregeljPRB}
Only 0.5\,K below, at $T_{N2}$\,=\,10.5\,K, 
the second transition to a predominantly 
 amplitude modulated magnetic ground state with the
 incommensurate (ICM) magnetic 
vector {\bf q$_{ICM}$}=($\frac{1}{2}$\,0.463\,0) occurs\cite{PregeljPRL}
 and is accompanied by a
 spontaneous electric polarization pointing
 perpendicular to the magnetic moments and to {\bf q$_{ICM}$}.
The magnetic susceptibility was explained originally by assuming dominant 
interactions within tetramers.\cite{Becker}
However, the tetramer model cannot explain the observed ICM
magnetic structure, essential for the ME effect in this system. 
The ambiguity obviously stems from structural complexity and therefore calls for additional experimental and theoretical investigations.

In this article, we disclose the underlying microscopic model for FeTe$_2$O$_5$Br
by a combination of density functional theory (DFT) calculations,
many-body model considerations, magnetization and
 electron spin resonance measurements. Our 
DFT results  surprisingly
reveal that long Fe-O-Te-O-Fe super-superexchange bridges are stronger
than  some shorter and more direct Fe-O-Fe ones. Consequently,
a quasi-one-dimensional model of alternating AFM Fe chains weakly
 coupled by frustrated two-dimensional interactions has been derived and 
verified by experiments.
In this picture, the ICM magnetic structure appears naturally.
It also reveals a direct effect of the interchain Fe-O-Te-O-Fe exchange pathways on the ICM component of the magnetic vector along the crystallographic $b$-axis, $q^y_{ICM}$. Close resemblance of its temperature dependence{ \cite{PregeljPRB}} to that of the electric polarization\cite{PregeljPRL} emphasizes the importance of the exchange striction on the interchain bonds that host easily polarizable Te$^{4+}$ lone-pair electrons, thus revealing the microscopic origin of the ME effect.
A similar mechanism is expected to be active also in selenite chain compounds.

\section{Experimental Details}

High-quality single crystals of FeTe$_2$O$_5$Br were grown by the standard chemical-vapor-phase method, reported elsewhere.\cite{Becker} Their crystallinity and phase purity have been confirmed by x-ray diffraction, which showed no sign of potential impurity phases. 

Magnetization and electron spin resonance (ESR) measurements were performed in pulsed magnetic fields up to 30\,T in the temperature range between $1.5$\,K and 20\,K at the Institute for Materials Research, Tohoku University, Sendai, Japan. The ESR experiment was conducted using non-polarized microwave radiation at fixed resonant frequencies between 95\,GHz and 405\,GHz.\cite{NojiriEPR} 

Single crystal x-ray synchrotron diffraction data were acquired at the BM01A Swiss-Norwegian
Beamline of ESRF (Grenoble, France). Data sets ($\sim$780 reflections per temperature
point) were collected in the temperature range between 4.5\,K and 35\,K at a wavelength of 0.64\,\AA,
using a closed-cycle He cryostat mounted on a six-circle kappa diffractometer KUMA, while the interatomic
distances were refined using the SHELXL97 program.\cite{SHELXL}

\section{Results}

\subsection{ Density Functional Theory}

\begin{table} [t!]

\caption{Exchange, $J_i$, and easy-plane anisotropy, $D$, parameters given in
units of Kelvin and normalized by the dominant $J_2$, as calculated by DFT calculations for
GGA+$U$,\,$U$=0, 3\,eV, and 5\,eV.
The last column corresponds to the parameters obtained  from magnetic ground state minimization and used for the AFMR simulation (AFMR + GS min). 
\label{tab1}}
\begin{ruledtabular}
\begin{tabular}{c|cc|cc|cc|cc}

$J_i$  & \multicolumn{2}{c|} {\bf GGA}   & \multicolumn{2}{c|}{\bf GGA+U}  & \multicolumn{2}{c|}{\bf GGA+U}  & \multicolumn{2}{c}{\bf AFMR +}  \\
       & \multicolumn{2}{c|}{} &  \multicolumn{2}{c|}{\bf\boldmath{$U=3$}eV}  & \multicolumn{2}{c|}{\bf\boldmath{$U=5$}eV}  & \multicolumn{2}{c}{\bf GS min} 
        \\ \hline
      &    (K)      & ($J_2$) &    (K)      & ($J_2$) &       (K)   & ($J_2$) &     (K)     & ($J_2$) \\
\hline                             
$J_1$  &   30.5      &   0.60      &   13.4      &   0.46      &  6.86       &   0.35      & 8.9    &   0.47      \\
$J_2$  &   51.1      &   1         &   29.2      &   1         &  19.6       &   1         & 19.0      &   1          \\
$J_3$  &   17.3      &   0.34      &    9.7      &   0.33      &  6.66       &   0.34      & 6.2       &   0.324       \\
$J_4$  &   35.3      &   0.69      &   18.1      &   0.62      & 11.56       &   0.59      & 11.8      &   0.62       \\
$J_5$  &    4.0      &   0.078     &    1.2      &   0.042     & -0.02       &   -0.001    & 0.8      &   0.042     \\
$J_6$  &   15.5      &   0.30      &    7.9      &   0.27      &  5.10       &   0.26      & 5.0       &   0.265       \\
$D$    &      -      &    -        &      -      &    -        &     -       &    -        & 0.18      &   0.0095    \\

\end{tabular}
\end{ruledtabular}
\end{table}

The underlying Heisenberg Hamiltonian parameters 
for FeTe$_2$O$_5$Br (Table~\ref{tab1}) were determined by 
total energy calculations. A full potential local orbital basis set~\cite{FPLO}
 and generalized gradient approximation (GGA)
 as well as GGA+$U$ functionals were used. The Hubbard parameter
 $U$ was chosen as 3 and 5\,eV  to take into account the intra-atomic Coulomb interactions. 
The obtained network of exchange interactions (Fig.~\ref{figDFT}) 
was checked for consistency by
N$^{th}$-order muffin-tin orbital downfolding~\cite{NMTO} calculations. 

\begin{figure*}  [t!]
\includegraphics[angle=0,width=0.7\textwidth]{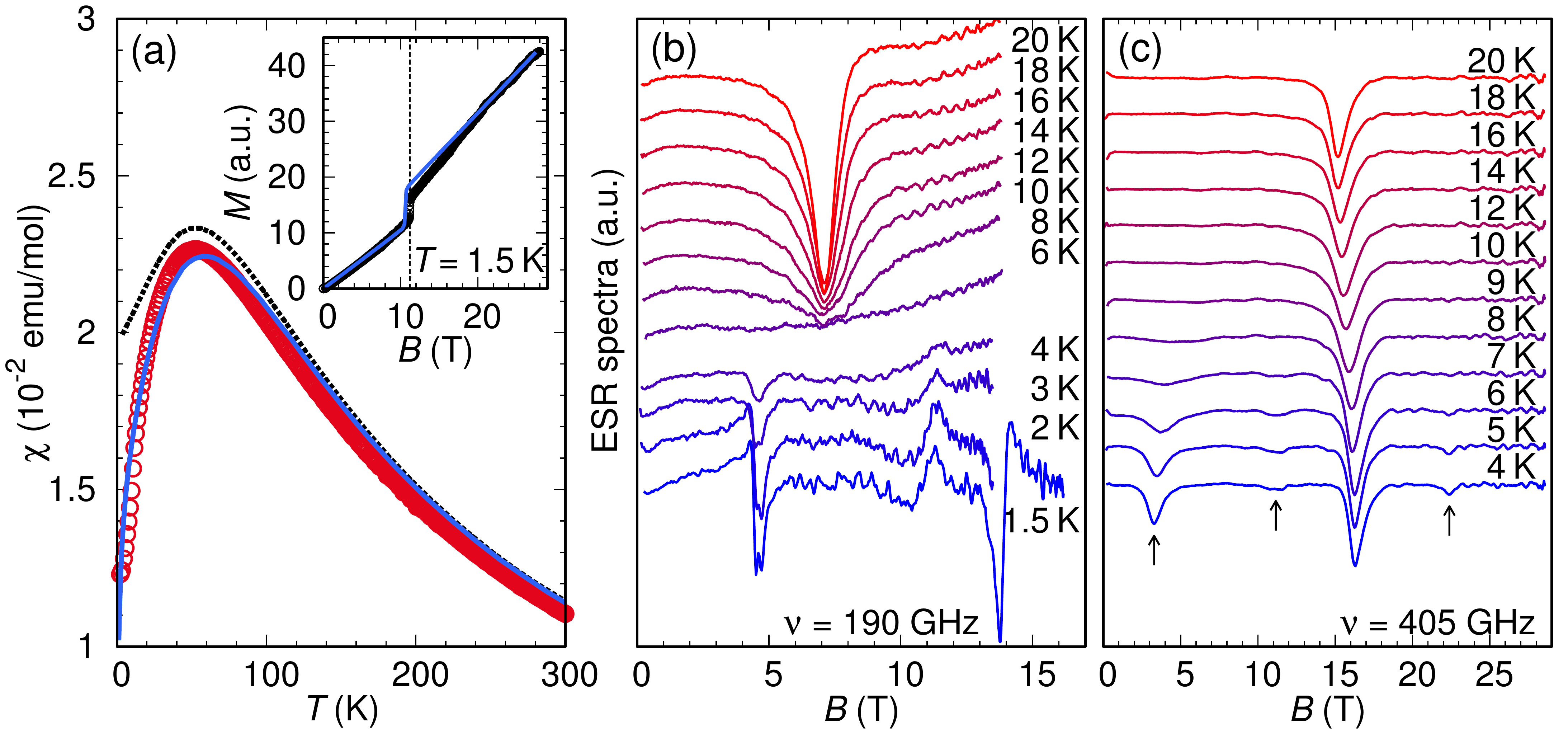}
\caption{(Color online)  (a) Measured magnetic susceptibility (circles) and comparison to the quantum Monte
Carlo (solid line) and the classical (dashed line) calculations for an alternating $S$\,=\,5/2 chain with $J_4/J_2$\,=\,0.6 and $J_2$\,=\,16\,K. Inset: High-field magnetization measured at $T$\,=\,1.5\,K (circles) and the simulation to the model described in the text (solid line). The dashed line indicates the position of $B_{SF}$. (b) Temperature evolution of ESR spectra at 190\,GHz and (c) 405\,GHz for $B\,||\,b$. Arrows indicate antiferromagnetic resonance modes emerging below the N\'{e}el transition temperature.}
\label{fig1}
\end{figure*}

The six considered exchange couplings are all antiferromagnetic. 
As expected, the intra\-cluster (within a tetramer)
 Fe-O-Fe exchange $J_2$ is the strongest. Strikingly,
 the second strongest exchange is not the remaining
 intracluster Fe-O-Fe exchange $J_1$, 
but rather the intercluster super-superexchange 
interaction $J_4$,  mediated through the Fe-O-Te-O-Fe bridges.
This result discloses that in tellurite oxohalides the super-superexchange interactions through long bridges involving Te$^{4+}$ ions are
important and should not be neglected on the basis of simplified structural arguments.
We advise similar caution also in case of Se$^{4+}$ ions in selenites.\cite{Janson, Janson2}
The two dominant couplings, $J_2$ and $J_4$,
 thus effectively form alternating Fe$^{3+}$ spin chains. 
The rest of the considered exchange interactions are weaker and provide
 frustrated interchain interactions as shown in Fig.~\ref{figDFT}. We  note that the DFT results leave some freedom concerning the overall energy scale, but the ratios of $J_i$'s are
expected to be subject only to small errors.\cite{azurite} However, the relative sizes of $J_i$'s are  somewhat dependent on the
 chosen value of $U$, with $J_1$ and $J_5$ showing the highest sensitivity. Thus, 
we suspect that they could be modified by  very small lattice distortions anticipated at the ME transition.

\subsection{Magnetic properties}

\label{sec:M}

We now consider the simple Heisenberg model,
\begin{equation}
H = \sum_{\langle i,j \rangle }J_{\langle i,j \rangle} \vec{S}_i \cdot \vec{S}_j,
\label{eq:Hop}
\end{equation}
First, we retain only the leading two exchange constants
$J_2$ and $J_4$, forming an alternating chain.
Quantum Monte Carlo simulations of the bulk magnetic susceptibility
for alternating $S=5/2$ chains with $L=60$ sites 
 were carried out with the ALPS 1.3 \cite{alps} directed loop application \cite{alps-sse} in the stochastic series
expansion framework.\cite{Sandvik} For $J_2$\,=\,16\,K and the
DFT determined ratio $J_4/J_2$\,=\,0.6  we find a very good agreement between
our simulations and the experimental data for $B$\,=\,0.1\,T and $B||b$ [Fig.~\ref{fig1}(a)].
A rather good agreement is also achieved 
when classical spins are considered (see appendix \ref{sec:appA} for further details). 
However, since the tetramer model can also fit the susceptibility
data,\cite{Becker} we need an independent experimental proof to
discriminate between the two models and to fix the precise values of $J$'s.

 Additional magnetization measurements performed 
at $T=1.5$\,K for $B$ parallel to the ICM direction ($b$ axis),
show a linear increase of the magnetization
with $B$ and the existence of a clean step  at around $B_{SF}=11$\,T 
[inset in Fig.~\ref{fig1}(a)].
Such a behavior is reminiscent of a spin-flop-like process, implying the presence of considerable magnetic anisotropy  which was not considered in the model
calculations.

  Finally, we note that no anomaly has been observed around 5\,T where the peak in the temperature dependence of the dielectric constant disappears.\cite{PregeljPRB} This suggests that the multiferroic ground state
 is not broken until $B_{SF}$ is reached and that the observed dielectric response probably reflects saturation of the ferroelectric domains by ME coupling.

\subsection{Electron spin resonance}

Magnetic anisotropies can be determined by  ESR,\cite{PregeljNi5, Herak, ZorkoPRL08, ZorkoPRL11} 
 which in the magnetically ordered state allows a detection of collective
 low-energy magnetic excitations at {\bf Q}\,=\,0,\,$\pm$\,{\bf q}$_{ICM}$ -- the
 so-called antiferromagnetic resonance (AFMR) modes.\cite{Huvonen}

In the paramagnetic state, i.e., far above $T_{N1}$ (at 300\,K), a strong ESR signal at $g$\,=\,2.005(5) is
observed, as expected for Fe$^{3+}$ ($S$\,=\,5/2)  ions.  At $T_{N1}$ the paramagnetic signal disappears and
is replaced by new resonances [Fig.~\ref{fig1}(b) and (c)] shifted with respect to the paramagnetic value. 
The new resonances below $T_{N1}$ are attributed to AFMR modes.
We note that at 405\,GHz one of the AFMR modes appears relatively close to the paramagnetic resonance,
screening its disappearance and causing a shift of the resonant line (Fig.~\ref{fig1}c).

\begin{figure} [t!]

\includegraphics[angle=0,width=0.47\textwidth]{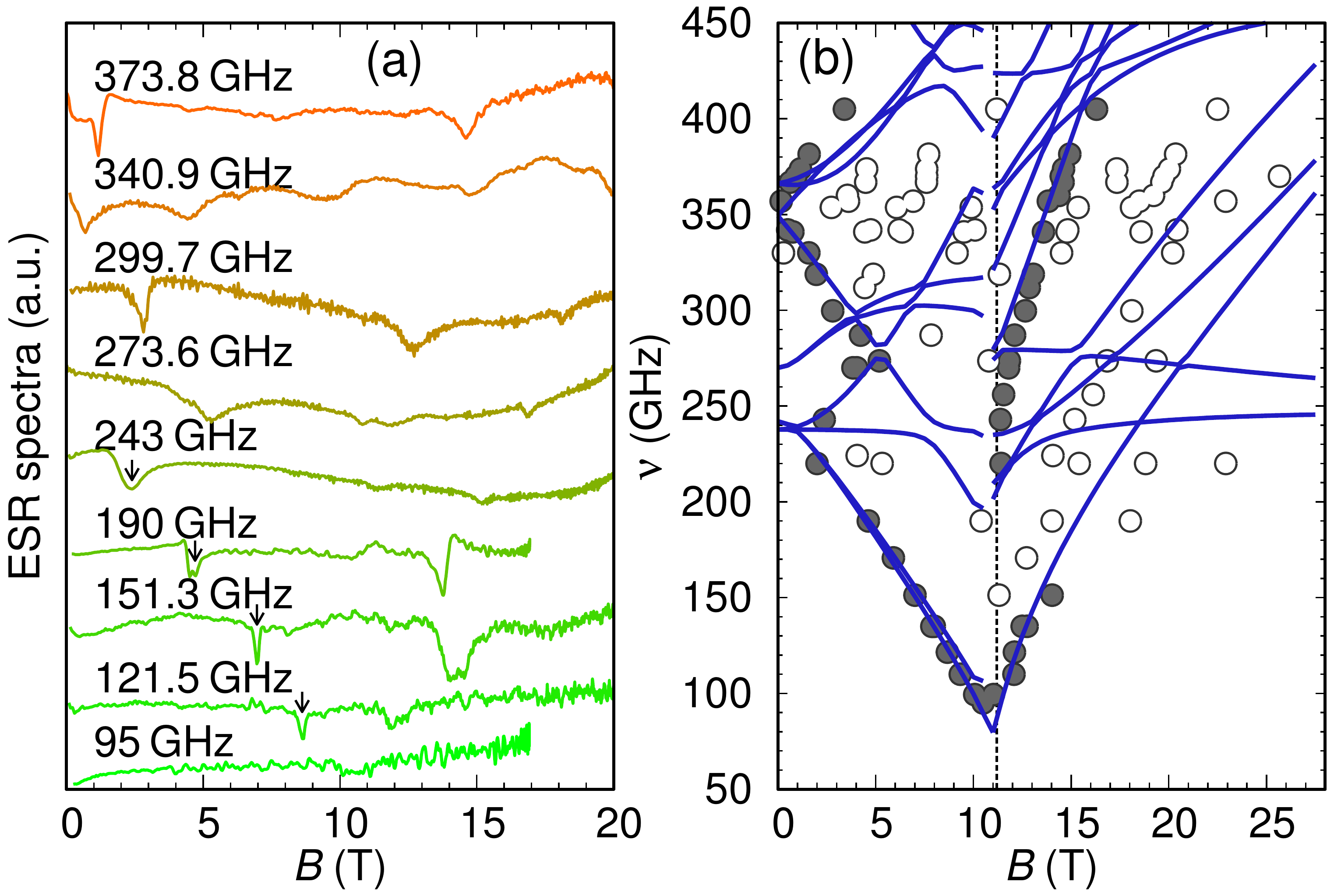}

\caption{(Color online)  (a) Field dependence of AFMR spectra at 1.5\,K for $B||b$ and various resonant frequencies between 95 and 405\,GHz. (b) Frequency-field dependence for intense (solid circles) and weak (open circles) AFMR modes. Solid lines are simulations to the model described in the text. The dashed line indicates $B_{SF}$.}

\label{fig2}

\end{figure}

 Since resonant fields for individual AFMR lines strongly depend on the resonance frequency [Fig.~\ref{fig2}(a)],  we were able to derive their frequency-field dispersions in the range 95 to 405 GHz and 0 to 30 T at 2\,K [Fig.\,\ref{fig2}(b)].
The lowest and at the same time the most intense excitation mode marked by arrows in Fig.~\ref{fig2}(a) shows
 a zero-field gap $\Delta\nu_{ZF}$\,$\sim$\,240\,GHz  [see Fig.~\ref{fig2}(b)]
corroborating a sizable magnetic anisotropy and in agreement with high-field magnetization 
measurements.  With increasing magnetic field the gap reduces to $\sim$\,100\,GHz at $B_{SF}$,
 where the slope is reversed and the gap increases again with increasing field. In addition, at least five more high-frequency modes were detected [Figs.~\ref{fig1}(b) and \ref{fig2}(a)], which also dramatically change their behavior at $B_{SF}$.

\subsection{Magnetic Ground State and AFMR Simulations}

In order to relate the high-field magnetization and the AFMR results to DFT calculations the first step 
is to determine the magnetic ground state of the infinite spin lattice.

In view of the hard axis anisotropy we simplify
the  Heisenberg model Eq.~(\ref{eq:Hop}) further and
consider in what follows coplanar moments.
 In this case the classical energy can be written as a function of the angle $\theta_i$ corresponding to the orientation of the spin $i$ with respect to the predetermined direction within the confined plane
\beqa
E_{Cl}&=&S^2\sum_{\langle i,j \rangle}J_{\langle i,j \rangle} \cos(\theta_j-\theta_i).
\eneqa
The classical energy thus depends only on the differences of orientations from spin to spin.
 Since Fe$^{3+}$ ions are in the high-spin $S$\,=\,5/2 state,  this classical
treatment of the model is reasonably justified.

For a spiral order, $E_{Cl}$ can be written as a function of  8 angles $\theta_i$  and the twist angles $q^y$ and $q^z$ defined as follows:
\beqa
\theta_{i,(n+1)r_y+mr_z} &=& \theta_{i,nr_y+mr_z}+q^y  \nn\\
\theta_{i,nr_y+(m+1)r_z} &=& \theta_{i,nr_y+mr_z}+q^z, 
\eneqa 
where $n$ and $m$ are the cell indices.

We used the conjugate gradient method to minimize the energy and to find the magnetic structure of the classical ground state. In all  our calculations we 
fixed the sizes of the magnetic moments and neglected their amplitude modulation.\cite{PregeljPRL}

Starting from the parameters obtained from the GGA+$U$ calculations with $U=0$~eV, 3~eV, and 5~eV  (Table~\ref{tab1}), the solution immediately converges to an ICM order.
For $U=3$~eV the experimental $q^y_{ICM}=0.463$ is reached by fine tuning of the interchain couplings, i.e.,
$J_1$, $J_3$, and $J_6$, up to 2\,\% (Table~\ref{tab1}).
We note that using the tetramer model parameters, i.e., $|J_1|=|J_2|\sim|J_3|\gg|J_4|$, $|J_5|$, $|J_6|$,\cite{Becker} we were unable to generate the ICM magnetic order, which clearly demonstrates that the tetramer model is inappropriate.

Due to the excellent agreement with $q^y_{ICM}$ we now use the optimized parameters
 (Table~\ref{tab1}, fourth column) to calculate the AFMR modes.
 For computing the equations of motion \cite{PregeljNi5}
 we define a finite magnetic lattice comprising 7 unit cells
 coupled along the crystallographic $b$ axis, i.e., mimicking the experimentally observed $q_{ICM}^y$\,=\,0.463\,$\sim$\,3/7. We assume 5\,$\mu_B$ for the size of Fe$^{3+}$ moments and introduce an easy plane anisotropy, $D S_z^2$. Here $D$ is the magnitude of single-ion anisotropy.
To fix the direction of the hard axis along $z$\,=\,(0.31,\,0,\,0.95) we take into the account the orientation of the ordered magnetic moments \cite{PregeljPRL} and a two-fold screw axis, which coincides  with the $b$ axis.
The only free parameters left in these calculations are thus $D$ and the strength of the $J_2$ interaction (all other interactions are scaled appropriately).

The strongest AFMR mode is best described [Fig.~\ref{fig2}(b)] by $J_2$\,=\,19.0\,K and $D$\,=\,0.18\,K
(Table~\ref{tab1}).
We stress that the theoretical curve nicely reproduces the softening of this mode up to $B_{SF}$ and its reversed character at higher fields.
At the same time, the dispersions of all other
 intense higher-frequency AFMR modes are 
in convincing agreement with the experiment [Fig.~\ref{fig2}(b)]. 
 We note that the calculated zero-field magnetic ground state is a coplanar cycloidal state, whereas the large number of calculated modes corresponds to 56 sublattice magnetizations considered in the calculations.
Moreover, the same set of parameters
 perfectly simulates the magnetization response to the
 applied magnetic field including the magnetization step at $B_{SF}$ [inset in Fig.~\ref{fig1}(a)].
To conclude this part, we emphasize a remarkable agreement between experiments and ratios of $J$'s obtained by the DFT calculations for GGA+$U$ with $U$\,=\,3\,eV achieved only by scaling slightly the exchange interactions.

\section{Discussion}

\begin{figure}[t!]
\includegraphics[width=0.42\textwidth]{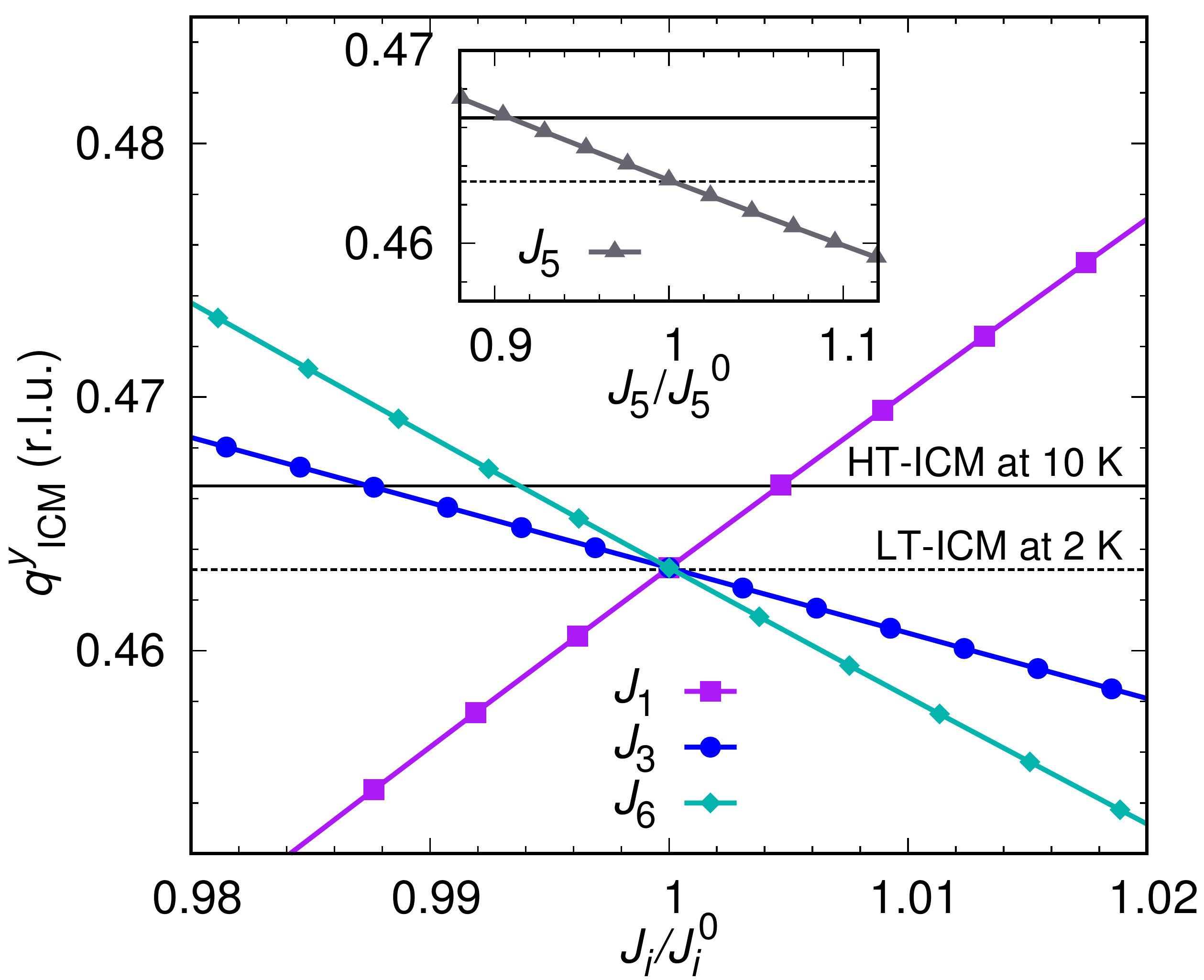}
\caption{(Color online) Dependence of $q^y_{ICM}$ on interchain exchange coupling constants ratios, $J_i/J_i^{0}$ for $i$\,=\,1,\,3,\,6, calculated by direct minimization of the classical energy, where $J_i^{0}$ are exchange parameters used in AFMR simulations. Inset: the same dependence calculated for $J_5$. }
\label{figQicJi}
\end{figure}

The main result of this study is the discovery that 
 in FeTe$_2$O$_5$Br some Fe-O-Te-O-Fe exchange pathways are stronger than some shorter Fe-O-Fe ones. Thus, the system has to be treated as a system of alternating $S$\,=\,5/2 chains weakly coupled by frustrated interactions, in contrast to the previously proposed tetramer model.
The knowledge of the appropriate spin model allows us to investigate 
the ME mechanism from a microscopic perspective. 
In order to identify the exchange pathway responsible for the ME effect we first recall
 that $q^y_{ICM}$ is temperature independent in the high-temperature ICM phase 
(between $T_{N1}$ and $T_{N2}$), while in the low-temperature ICM phase 
(below $T_{N2}$) it scales similarly to the electric polarization, i.e., it 
 behaves as ($T_{N2}-T$)$^{0.35}$.\cite{PregeljPRB}
To address this important point we return to the minimization of the classical energy for the infinite lattice and calculate how $q^y_{ICM}$ is affected by
 small changes of different $J_i$'s. 
Since the ICM modulation is perpendicular to the alternating chains,
it suffices to vary only the interchain exchange couplings $J_i$ ($i$=1,3,5,6) with respect to the
values that reproduce the AFMR modes at $T$\,=\,2\,K. 
In order to reproduce the decrease of $q^y_{ICM}$ observed below $T_{N2}$,
 we find that $J_1$ has to be reduced, while $J_3$, $J_5$, and $J_6$ have to be increased (Fig.~\ref{figQicJi}). 

\begin{figure}[t!]
\includegraphics[width=0.42\textwidth, angle=0, trim=0 0 0 10, clip=true]{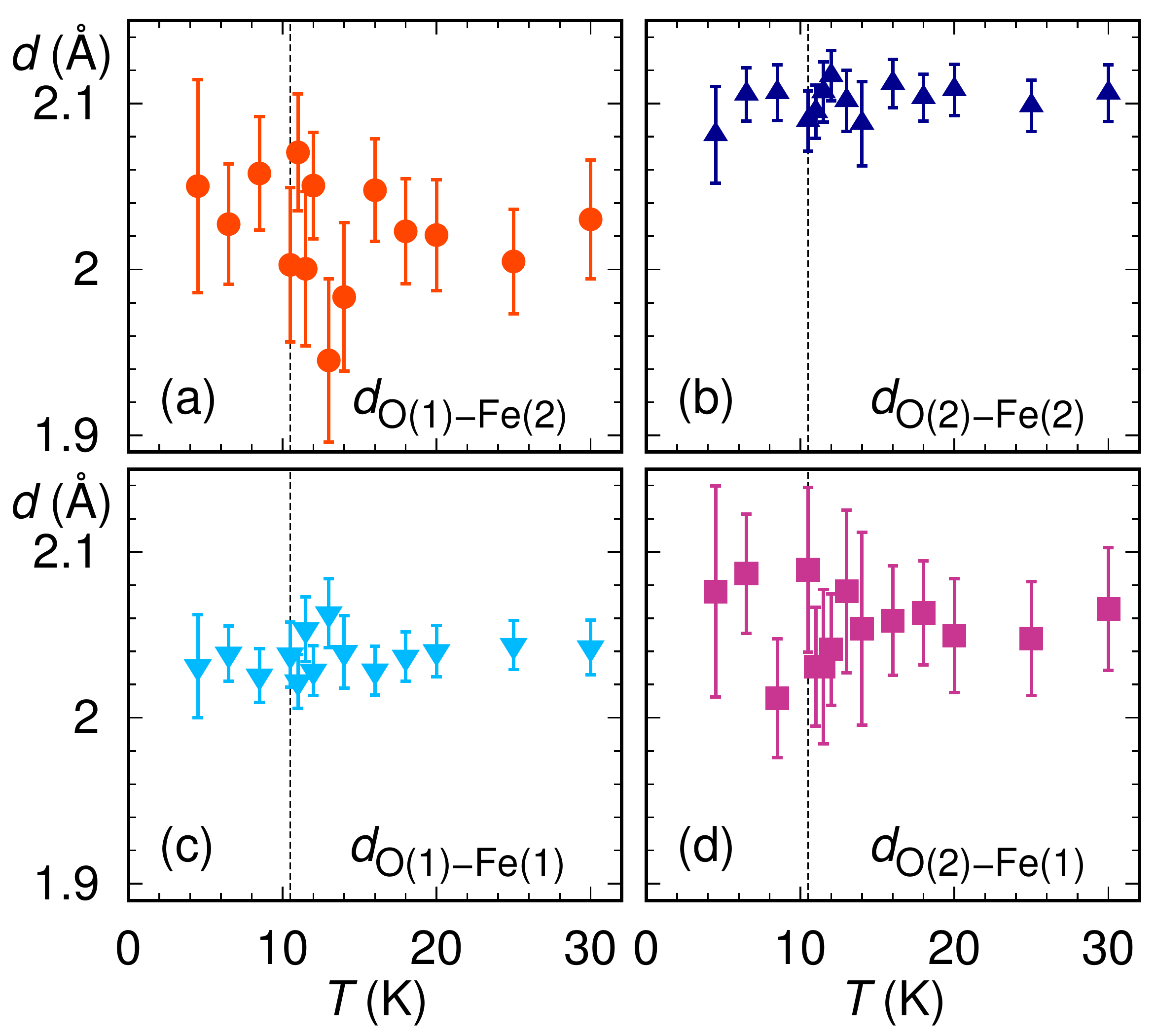}
\caption{Refinement results for the O-Fe distances obtained from synchrotron x-ray diffraction, corresponding to $J_1$ (a-d) and $J_3$ (d) exchange pathways.}
\label{O-Te}
\end{figure}

Given that $J_1$ and $J_3$ are superexchange interactions via Fe-O-Fe bridges, 
they are expected to behave according to Goodenough-Kanamori 
rules.\cite{Goodenough, Kanamori} If $J_1$ is responsible for 
the ME effect, then the Fe-O-Fe angle should be reduced or the Fe-O distance increased in 
order to weaken $J_1$. No significant structural changes
for this bond have been observed in our 
high-resolution synchrotron x-ray diffraction
 experiments  (Fig.~\ref{O-Te}) thus ruling out this possibility. An analogous conclusion
 can be derived also for $J_3$ [Fig.~\ref{O-Te}(d)]. We are thus 
left with the two remaining candidates -- $J_5$ and $J_6$ super-superexchange
 interactions -- which both involve Te$^{4+}$ ions.
Since the only sizable change at the ME transition
 corresponds to the shortening of the Fe$_2$-Te$_3$ distance, \cite{PregeljPRL} involved in the $J_5$ pathway, we assign the microscopic origin of the ME coupling to the changes of this interaction.
While the realization that the magnetic ordering breaks the inversion symmetry is essential to understand phenomenological aspects of the ME effect,\cite{PregeljPRL} the key to understanding the ME coupling at the microscopic scale lies in the low symmetry of the long Fe-O-Te-O-Fe bridges, which allows emergent magnetic order to provoke a softening of an appropriate phonon mode and hence a net electric polarization. We stress that this is possible even if the spin-orbit coupling is very small. 
This resembles Ni$_3$V$_2$O$_8$, where the exchange striction mechanism is proposed to be responsible for the ME couping.\cite{Harris}
Finally, we point out that tellurite oxohalide systems seem particularly inclined to such effects, since their long low-symmetry exchange bridges involve easily polarizable Te$^{4+}$ lone-pair electrons, which enable that already minimal changes in the strength of the exchange interactions are sufficient to produce a measurable magnetoelectric effect.

\section{Conclusions}

We have shown that the magnetic exchange 
network of the FeTe$_2$O$_5$Br system should be regarded as a
 system of alternating Fe chains with weaker frustrated interchain interactions.
Frustration is found to be responsible for the observed low-symmetry ICM magnetic 
ordering, essential for the establishment of the multiferroic phase.
 Additionally, we show that in FeTe$_2$O$_5$Br the ME effect 
 on the microscopic level originates from the exchange striction
 of interchain $J_5$ Fe-O-Te-O-Fe exchange pathway, where the net 
electric polarization comes from the Te$^{4+}$ lone-pair electrons. 
Finally, our findings clearly demonstrate that in tellurite
and selenite oxohalides 
one needs to be extremely cautious when making assumptions about
the exchange network, as  
arguments based solely on the crystal structure may be very misleading.
Long super-superexchange bridges 
can lead to surprisingly strong interactions
and their structural volatility is at the core of the ME effect.


\acknowledgments
This work has been supported by the
Swiss National Science Foundation project 200021-129899,
the DFG (SFB/TR~49 and HO~2325/4-2), and
by the Helmholtz Association through HA216/EMMI, and the Slovenian research agency (J1-2118).
The IMR group Tohoku acknowledges the support by GCOE: Weaving Science Web beyond Particle-Matter Hierarchy and Kakenhi from MEXT Japan. We acknowledge help from the ESRF Swiss-Norwegian Beamline scientists D. Chernyshov and Ya. Filinchuk with the x-ray experiment and data analysis.

\appendix

\section{Magnetic susceptibility calculations}

\label{sec:appA}

\begin{figure}[t!]
\centering
\includegraphics[width=0.42\textwidth, trim=0 0 0 0, clip=true]{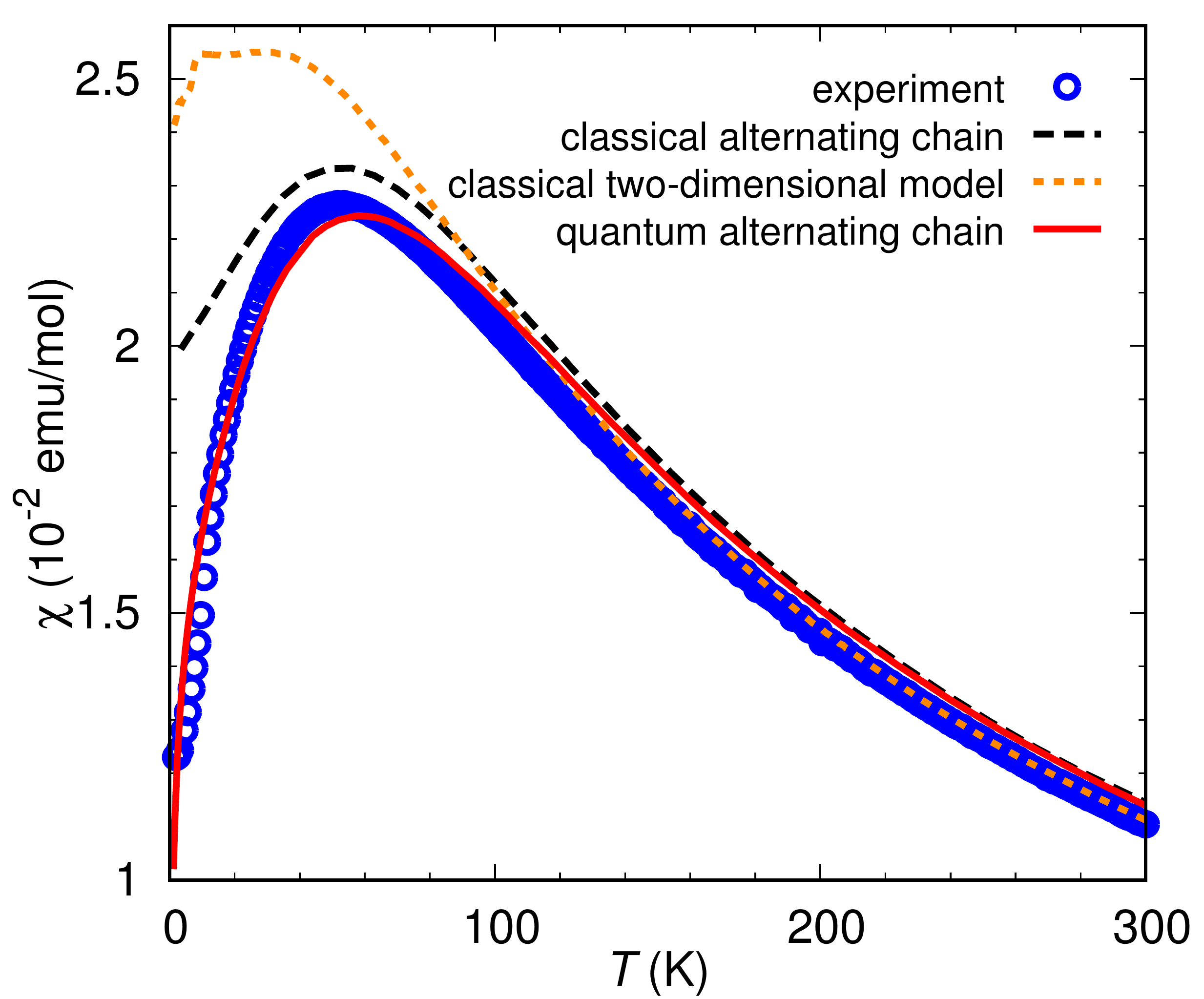}
\caption{(Color online)  Measured magnetic susceptibility (circles), classical (dashed line), and quantum (solid line) Monte Carlo results for an alternating $S=5/2$ chain
with 60 spins, $J_4/J_2=0.6$, and $J_2 = 16$~K.
Here we also include results for a two-dimensional $64\times64$ lattice
of classical spins (dotted line), using the exchange ratios
determined from the ground state minimization and AFMR calculations,
but with $D=0$ and scaled to $J_2 = 12$~K.
Statistical errors of the Monte Carlo sampling procedure are well below the width of the lines.}
\label{figSuscep}
\end{figure}

The magnetic susceptibility was computed by classical and quantum Monte Carlo
simulations of the Heisenberg model (\ref{eq:Hop}) and compared to our single crystal $\chi(T)$ measurements for $B$\,=\,0.1\,T and $B||b$. Results for the
alternating spin chain model were already presented in Sec.\ \ref{sec:M};
here we also present results for magnetic susceptibility of the
$64\times64$ lattice of classical spins with the exchange 
ratios determined from the ground state minimization and AFMR calculations (Fig.~\ref{figSuscep}). In order to explain the observed high-temperature 
behavior within the two-dimensional model, we need to scale to
$J_2 = 12$~K, while for simplicity we set $D=0$.
All numerical results are for a fixed system size
and periodic boundary conditions,
but we have checked that finite-size effects are negligible except possibly
at very low temperatures. Theoretical results were converted to experimental
units assuming a spectroscopic g-factor $g=2$.

Inspired by the scaling of the high-temperature behavior with spin
quantum number $S$, we used a phenomenological scaling
\begin{equation}
\chi_{S,\text{phen}}(T) = \chi_{\text{cl}}[S(S+1)\,T]
\label{eq:chi}
\end{equation}
to map the classical result $\chi_{\text{cl}}$ to the
quantum result for spin $S$. As is illustrated by the results for
the alternating chain, this works well at
high temperatures, but misses a suppression of $\chi$ by
quantum fluctuations at low temperatures.

In the two-dimensional model, interchain coupling is frustrated.
Since this gives rise to a sign problem in the quantum Monte Carlo
approach, we have to rely exclusively on the classical Monte
Carlo simulations for this case. The two-dimensional model fits the
 experimental magnetic susceptibility at high temperatures
at least as well as the chain model,  while it does not appear to be so
good around and below the maximum of $\chi(T)$ (Fig.~\ref{figSuscep}).
 However, we have to keep in mind
that in analogy to the one-dimensional model, quantum fluctuations
are expected to reduce $\chi(T)$ at low temperatures also in
two dimensions. Furthermore, the frustrated nature of the interchain
coupling might enhance this reduction. In any case, we attribute
the fact that a one-dimensional model yields a good effective description
to the frustration in the interchain coupling. Finally, we note
that the  two-dimensional model exhibits signatures
of an ordering transition at $T\approx9.5$~K. This is remarkably
close to the experimentally observed ordering transitions, although
the Mermin-Wagner theorem forbids a true finite-temperature
ordering transition in a strictly two-dimensional Heisenberg model.

In conclusion, the fact that  one-dimensional, two-dimensional,
and tetramer\cite{Becker} models provide good
fits to the magnetic susceptibility of FeTe$_2$O$_5$Br
demonstrates that further data is needed to clarify the
nature of the microscopic exchange network.


\begin{thebibliography}{000}

\bibitem{Lacroix} {\em Introduction to Frustrated Magnetism}, edited by C. Lacroix, P. Mendels, and F. Mila (Springer-Verlag, Berlin, 2011).

\bibitem{Cheong} S.-W. Cheong and M. Mostovoy, { Nature Mater.} {\bf 6}, 13 (2007).

\bibitem{Khomskii} D. Khomskii, { Physics} {\bf 2}, 20 (2009).

\bibitem{Eerenstein} W. Eerenstein, N. D. Mathur, and J. F. Scott, { Nature} {\bf 442}, 759 (2006).

\bibitem{Hur} N. Hur, S. Park, P. A. Sharma, J. S. Ahn, S. Guha, and S.-W.
Cheong, {Nature} {\bf 429}, 392 (2004).

\bibitem{Lawes} G. Lawes, A. P. Ramirez, C. M. Varma, and M. A. Subramanian,
  Phys. Rev. Lett. {\bf 91}, 257208 (2003).

\bibitem{Bos} J.-W. G. Bos, C, V. Colin, and T. T. M. Palstra,
  Phys. Rev. B {\bf 78}, 094416 (2008).

\bibitem{Zaharko} O. Zaharko, H. R\o nnow, J. Mesot, S. J. Crowe, P. J. Brown, A. Daoud-Aladine, A. Meents, A. Wagner, M. Prester, H. Berger, and D. McK. Paul, { Phys Rev. B} {\bf 73}, 064422 (2006). 

\bibitem{Hiroshi} H. Murakawa, Y. Onose, K. Ohgushi, S. Ishiwata, and Y. Tokura, { J. Phys. Soc. Jpn.} {\bf 77}, 043709 (2008).

\bibitem{PregeljPRL} M. Pregelj, O. Zaharko, A. Zorko, Z. Kutnjak, P. Jegli\v{c}, P. J. Brown, M. Jagodi\v{c}, Z. Jagli\v{c}i\'{c}, H. Berger, and D. Ar\v{c}on, { Phys. Rev. Lett.} {\bf 103}, 147202 (2009).

\bibitem{Becker} R. Becker, M. Johnsson, R. K. Kremer, H.-H. Klauss, and P. Lemmens, { J. Am. Chem. Soc.} {\bf 128}, 15469 (2006).

\bibitem{Becker2} R. Becker and M. Johnsson, {J. Solid. State Chem.} {\bf 180}, 1750 (2007).

\bibitem{Janson} O. Janson, W. Schnelle, M. Schmidt, Yu. Prots, S.-L. Drechsler, S. K. Filatov, and H. Rosner, { New J. Phys.} {\bf 11}, 113034 (2009).

\bibitem{Janson2} O. Janson, A. A. Tsirlin, E. S. Osipova, P. S. Berdonosov, A. V. Olenev, V. A. Dolgikh, and
  H. Rosner, { Phys. Rev. B} {\bf 83}, 144423 (2011).

\bibitem{Deisenhofer} J. Deisenhofer, R. M. Eremina, A. Pimenov, T. Gavrilova, H. Berger, M. Johnsson, P. Lemmens, H.-A. Krug von Nidda, A. Loidl, K.-S. Lee, and M.-H. Whangbo,  {Phys Rev. B} {\bf  74}, 174421 (2006). 

\bibitem{Das} H. Das, T. Saha-Dasgupta, C. Gros, and R. Valent\'{i}, {Phys Rev. B} {\bf  77}, 224437 (2008).

\bibitem{Valenti} R. Valent\'{\i}, T. Saha-Dasgupta, C. Gros, and H. Rosner, {Phys Rev. B} {\bf  67}, 245110 (2003). 

\bibitem{Goodenough} J. B. Goodenough, { Phys. Rev.} {\bf 100}, 564 (1955).

\bibitem{Kanamori} J. Kanamori, { J. Phys. Chem. Solids} {\bf 10}, 87 (1959).

\bibitem{PregeljPRB} M. Pregelj, A. Zorko, O. Zaharko, Z. Kutnjak, M. Jagodi\v{c}, Z. Jagli\v{c}i\'{c}, H. Berger, M. de Souza, C. Balz, M. Lang, and D. Ar\v{c}on, { Phys. Rev. B} {\bf  82}, 144438 (2010).


\bibitem{NojiriEPR} H. Nojiri, M. Motokawa, K. Okuda, H. Kageyama, Y. Ueda and H. Tanaka, {J. Phys. Soc. Jpn. Suppl. B} {\bf 72}, 109 (2003).

\bibitem{SHELXL} G. M. Sheldrick, computer program SHELXL97, University of G\"{o}ttingen: G\"{o}ttingen Germany
1997.


\bibitem{FPLO} K. Koepernik and H. Eschrig,
  {Phys. Rev. B} {\bf 59}, 1743 (1999); {\tt
    http://www.FPLO.de}.

\bibitem{NMTO} O. K.  Andersen and  T. Saha-Dasgupta,
  {Phys. Rev. B} {\bf 62}, R16219 (2000).

\bibitem{azurite} H. O. Jeschke,  I. Opahle, H. Kandpal, R. Valent\'{i}, H. Das, T. Saha-Dasgupta, O. Janson, H. Rosner, A. Br\"{u}hl, B. Wolf, M. Lang, J. Richter, S. Hu, X. Wang, R. Peters, T. Pruschke, and A. Honecker, { Phys. Rev. Lett. } {\bf 106}, 217201 (2011).

\bibitem{alps} A. F. Albuquerque, F. Alet, P. Corboz, P. Dayal, A.
 Feiguin, S. Fuchs, L. Gamper, E. Gull, S. G\"urtler, A. Honecker, R.
Igarashi, M. K\"orner, A. Kozhevnikov, A. L\"auchli, S. R. Manmana, M.
 Matsumoto, I. P. McCulloch, F. Michel, R. M. Noack, G. Paw{\l}owski, L.
 Pollet, T. Pruschke, U. Schollw\"ock, S. Todo, S. Trebst, M. Troyer, P.
 Werner, and S. Wessel, 
J. Magn. Magn. Mater. {\bf 310}, 1187 (2007).

\bibitem{alps-sse} F. Alet, S. Wessel, and M. Troyer, Phys. Rev. E {\bf 71}, 036706 (2005).

\bibitem{Sandvik} O. F. Sylju{\aa}sen and A. W.\ Sandvik, Phys. Rev. E
 {\bf 66}, 046701 (2002).

\bibitem{PregeljNi5} M. Pregelj, A. Zorko, H. Berger, H. van Tol, L. C. Brunel, A. Ozarowski, S. Nellutla, Z. Jagli\v{c}i\'{c}, O. Zaharko, P. Tregenna-Piggott, and D. Ar\v{c}on, { Phys. Rev. B} {\bf  76}, 144408 (2007).

\bibitem{ZorkoPRL11} A. Zorko, M. Pregelj, A. Poto\v{c}nik, J. van Tol, A. Ozarowski, V. Simonet, P. Lejay, S. Petit, and R. Ballou, { Phys. Rev. Lett.} {\bf 107}, 257203 (2011).

\bibitem{Herak} M. Herak, A. Zorko, D. Ar\v{c}on, A. Poto\v{c}nik, M. Klanj\v{s}ek,
J. van Tol, A. Ozarowski, and H. Berger, { Phys. Rev. B} {\bf 84}. 184436 (2011).

\bibitem{ZorkoPRL08} A. Zorko, S. Nellutla, J. van Tol, L. C. Brunel, F. Bert,
F. Duc, J.-C. Trombe, M. A. de Vries, A. Harrison, and P. Mendels, { Phys. Rev.
Lett.} {\bf 101}, 026405 (2008).

\bibitem{Huvonen} D. H\"uvonen, U. Nagel, T. R\~{o}\~{o}m,
  Y. J. Choi, C. L. Zhang, S. Park, and S.-W. Cheong,
  { Phys. Rev.} B {\bf 80}, 100402(R) (2009).

\bibitem{Harris} A. B. Harris, T. Yildirim, A. Aharony, and O. Entin-Wohlman, {Phys. Rev. B} {\bf 73} 184433 (2006).





\end{thebibliography}
\end{document}